\documentclass[sigconf,authorversion]{acmart}

\AtBeginDocument{%
  \providecommand\BibTeX{{%
    \normalfont B\kern-0.5em{\scshape i\kern-0.25em b}\kern-0.8em\TeX}}}

\copyrightyear{2021}
\acmYear{2021}
\setcopyright{acmlicensed}\acmConference[AH2021]{12th Augmented Human International Conference}{May 27--28, 2021}{Geneva, Switzerland}
\acmBooktitle{12th Augmented Human International Conference (AH2021), May 27--28, 2021, Geneva, Switzerland}
\acmPrice{15.00}
\acmDOI{10.1145/3460881.3460933}
\acmISBN{978-1-4503-9030-9/21/05}

\acmSubmissionID{9}

\begin{document}

\title{xBalloon: Animated Objects with Balloon Plastic Actuator}

\author{Haoran Xie}
\affiliation{%
  \institution{Japan Advanced Institute of Science and Technology}
  \state{Ishikawa}
  \country{Japan}}
  \email{xie@jaist.ac.jp}
  
\author{Takuma Torii}
\affiliation{%
  \institution{Japan Advanced Institute of Science and Technology}
  \state{Ishikawa}
  \country{Japan}}
  
 \author{Aoshi Chiba}
\affiliation{%
  \institution{Japan Advanced Institute of Science and Technology}
  \state{Ishikawa}
  \country{Japan}}

  \author{Qiukai Qi}
\affiliation{%
  \institution{Japan Advanced Institute of Science and Technology}
  \state{Ishikawa}
  \country{Japan}}

\renewcommand{\shortauthors}{H. Xie, et al.}

\begin{abstract}
 Shape-changing interfaces are promising for users to change the physical properties of common objects. However, prevailing approaches of actuation devices require either professional equipment or materials that are not commonly accessible to non-professional users. In this work, we focus on the controllable soft actuators with inflatable structures because they are soft thus safe for human computer interaction. We propose a soft actuator design, called xBalloon, that is workable, inexpensive and easy-to-fabricate. It consists of daily materials including balloons and plastics and can realize bending actuation very effectively. For characterization, we fabricated xBalloon samples with different geometrical parameters and tested them regarding the bending performance and found the analytical model describing the relationship between the shape and the bending width. We then used xBalloons to animate a series of common objects and all can work satisfactorily. We further verified the user experience about the the fabrication and found that even those with no prior robotic knowledge can fabricate xBalloons with ease and confidence. Given all these advantages, we believe that xBalloon is an ideal platform for interaction design and entertainment applications.
\end{abstract}



\maketitle

\section{Introduction}
Shape-changing interfaces have emerged with the modifications on the real world objects in human computer interaction \cite{shape12}. Due to the direct interaction to users, the soft actuators have great potential in the transformable interfaces with soft materials with hand-on training on robotic skills \cite{softed14}. Often fabricated with deformable materials, soft actuators are compliant, thus secure and ideal for human computer interaction. One concern, however, that may constrain the applications of soft actuators for robotic education is the fabrication which usually requires professional apparatus and materials \cite{soft14}, such as 3D printing devices and silicone rubbers. To non-professional users without fabrication skills, these are neither readily accessible nor user-friendly enough. In order to generalize the their applications in common usage, soft actuators ideally need to be more available, affordable and easy-to-fabricate \cite{softpathahead16}.

\begin{figure}[t]
  \centering
  \includegraphics[width=\linewidth]{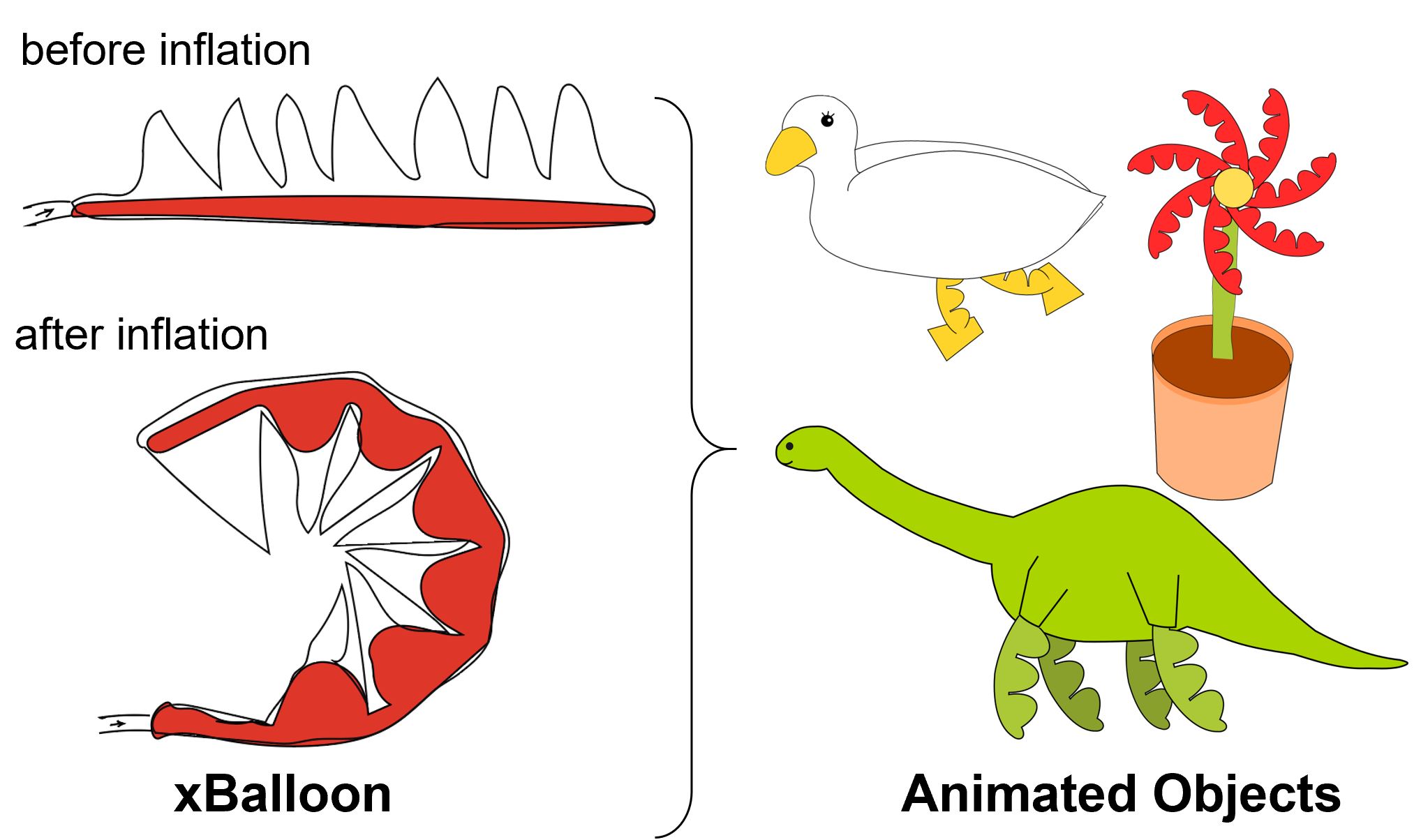}
  \caption{Animated objects with our proposed balloon plastic actuator, xBalloon.}
  \label{fig:concept}
\end{figure}

The goal of this study is to propose a novel mechanism for simple yet effective soft actuator that even those with no robotic knowledge can make easily and rapidly with daily materials and devices. Among the popular material options in this regard, we are particularly interested in the inflatable structures with balloons considering the ease of fabrication and availability of the material components. Many previous works have demonstrated the effectiveness of similar mechanisms in different applications, for example, art creation \cite{ba19} and shape construction \cite{ig07}. These examples use balloons to achieve linear deformation, while none can realize bending deformation, which can be useful in real world cases where rotary movement is required \cite{ni15}.

\begin{figure*}[t]
  \centering
  \includegraphics[width=\linewidth]{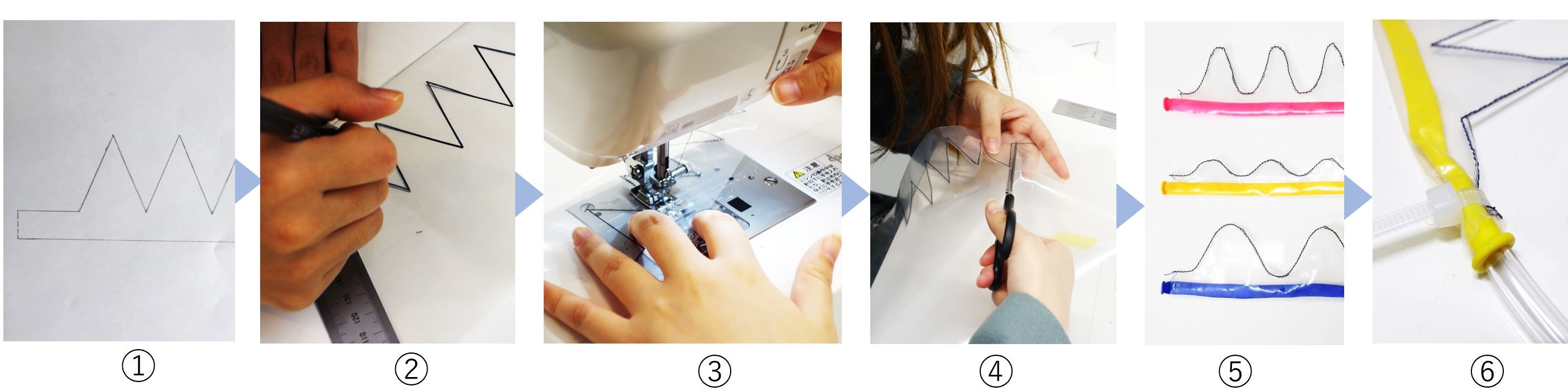}
  \caption{The fabrication procedure of xBalloon actuator includes six simple steps with the common and cheap materials.}
  \label{fig:fab}
\end{figure*}

In this study, we propose a balloon plastic actuator, xBalloon, composed of two inexpensive materials: a balloon and a polyethylene film. Here, the shape change is triggered by the expansion of the balloon inside the polyethylene film. The direction of deformation is controlled by covering the balloon with the polyethylene film. We evaluated the deforming performance and used this actuator to animate a series of objects for demonstration. Since this mechanism requires no professional skill and is inexpensive, we consider that xBalloon is suitable to be used for animated objects in interaction and entertainment applications, as illustrated in Figure~\ref{fig:concept}.

\begin{figure}[htbp]
  \centering
  \includegraphics[width=\linewidth]{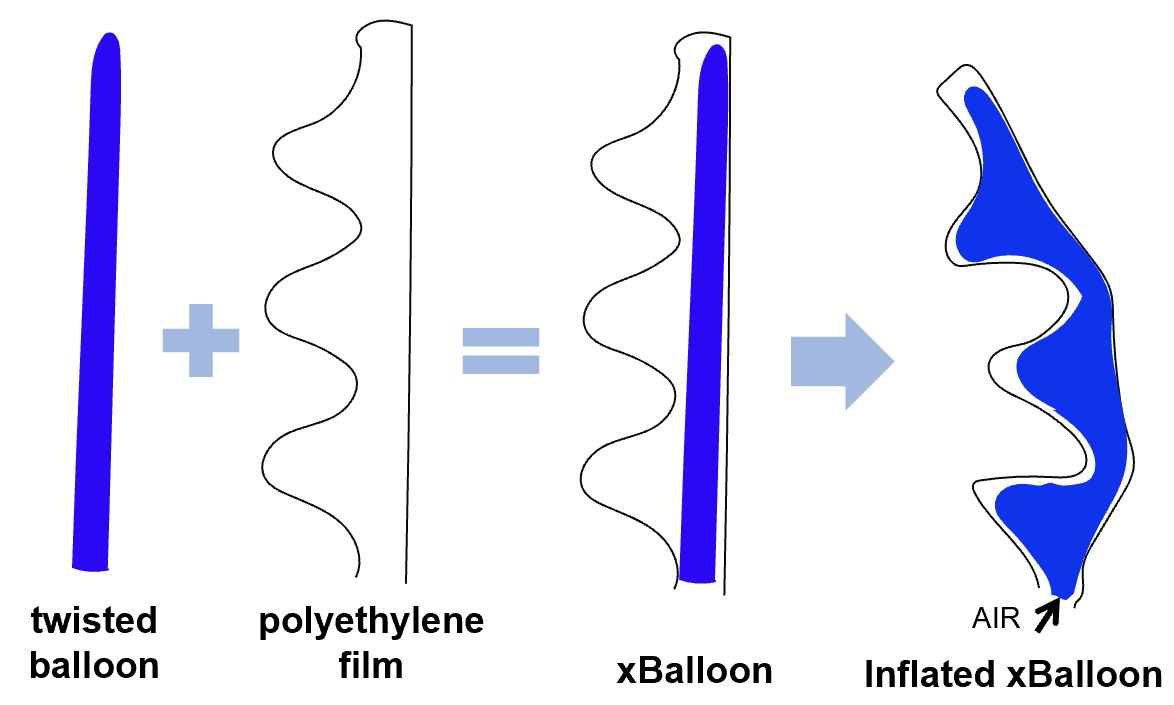}
  \caption{Component configuration of xBalloon actuator.}
  \label{fig:config}
\end{figure}

\section{Related work}
In the field of human augmentation, the usage of wearable devices is a common way to extend the human capabilities, such as supernumerary robotic limbs for balance support and joint load reduction~\cite{asada05}, head-mounted device for vision augmentation~\cite{ego20}, and tail-type joints for weight support and emotional expression \cite{xie19}. Recently, the wrist-mounted robotic arm was proposed for performing various daily tasks powered by magnetorheological clutches and hydrostatic transmission lines~\cite{robot20}. However, all these devices are rigid, thereby being dangerous in misoperation.

Soft actuators can keep the user safe with soft materials. The common soft actuators are made by 3D printed mold and resin like Soft Robotics Toolkit \cite{soft14}. Textile bending actuators utilized fabric with different stretch properties to achieve complex motions with embedded balloons inside \cite{soft18}. The sheet-like actuators were to support interaction with wearable garments and enable autonomous robotic systems \cite{zhu20}. The textile actuators were made for wearable robotic garments and applications of soft robotic \cite{soft21}. In this work, we especially focus on the balloon actuator with simple fabrication procedure.

Pneumatic balloon actuator was fabricated on a cantilever for actuating the flexible end-effector \cite{ba01}. PneUI presented pneumatically-actuated soft composite materials for shape-changing interactions \cite{yao13}. Sticky actuator was made by adhesive-backed inflatable pouches and used for rapid development of animated objects \cite{ni15}.  A soft inflatable joint was proposed using the blower-inflated structure of cylindrical links and joints with tension force \cite{kawa19}. BlowFab proposed an inflatable structure with laser cutting and blow molding techniques \cite{blow17}. Giacometti arm was fabricated by helium-filled balloons and pneumatic muscles \cite{suzumori17}. It is difficult for non-professional users to reproduce these soft actuators without professional skills and equipment. BPActuators proposed a lightweight and low-cost soft bending actuator with plastics and balloons \cite{qi2021bpactuators}. However, the exploration on object augmentation in dynamics representation with pneumatic actuators is absent for interaction purposes. In this work, we aim to propose a novel soft actuator that can augment the common objects to be animated and anyone can make easily.

\section{xBalloon}
xBalloon is made by balloon and polyethylene film as shown in Figure \ref{fig:config}. The polyethylene film is stitched along the rim sketches using a sewing machine, and the balloon is inserted into the film cavity. The xBalloon actuator can bend with inflated balloon inside the film cavity. As illustrated in Figure \ref{fig:config}, the length of left side of the actuator becomes shorter due to the air inflation, then the actuator begins to bend to the shorter side.

\subsection{Fabrication Procedure}
As illustrated in Figure \ref{fig:fab}, the specific procedures for fabrication of xBalloon actuator are detailed as below in six steps.  
\begin{enumerate} 
    \item Drawing a mold shape on a piece of paper;
    \item Placing the mold on a folded polyethylene film and tracing it with a marker pen;
    \item Sewing along the traced curves with the sewing machine;
    \item Cutting out the polyethylene film along the sewed curves; 
    \item Inserting a twist balloon into the cavity of the polyethylene film;
    \item Inserting the air tube of film cavity and tying the cavity mouth tightly with bundling bands.  
\end{enumerate}
Apart from the component materials in Figure \ref{fig:config}, a sewing machine, an air tube, a maker pen and a piece of plain paper are also required. All the materials needed here are very common and readily accessible with very low price, except the sewing machine. Note that instead of using a sewing machine, the stitched rims can also be generated manually. Here, the total cost of one xBalloon actuator is around 0.22 USD (United State Dollar), where balloon is 0.011 USD per unit and plastic is 0.21 USD. 

\subsection{Actuation Performance}
In order to verify the actuation performance of xBalloon actuator, we conducted the performance experiments to explore the relationship between the fabrication shape and bending width of xBalloon actuator.

\subsubsection*{Performance Experiment}
\begin{figure}[tb]
  \centering
  \includegraphics[width=\linewidth]{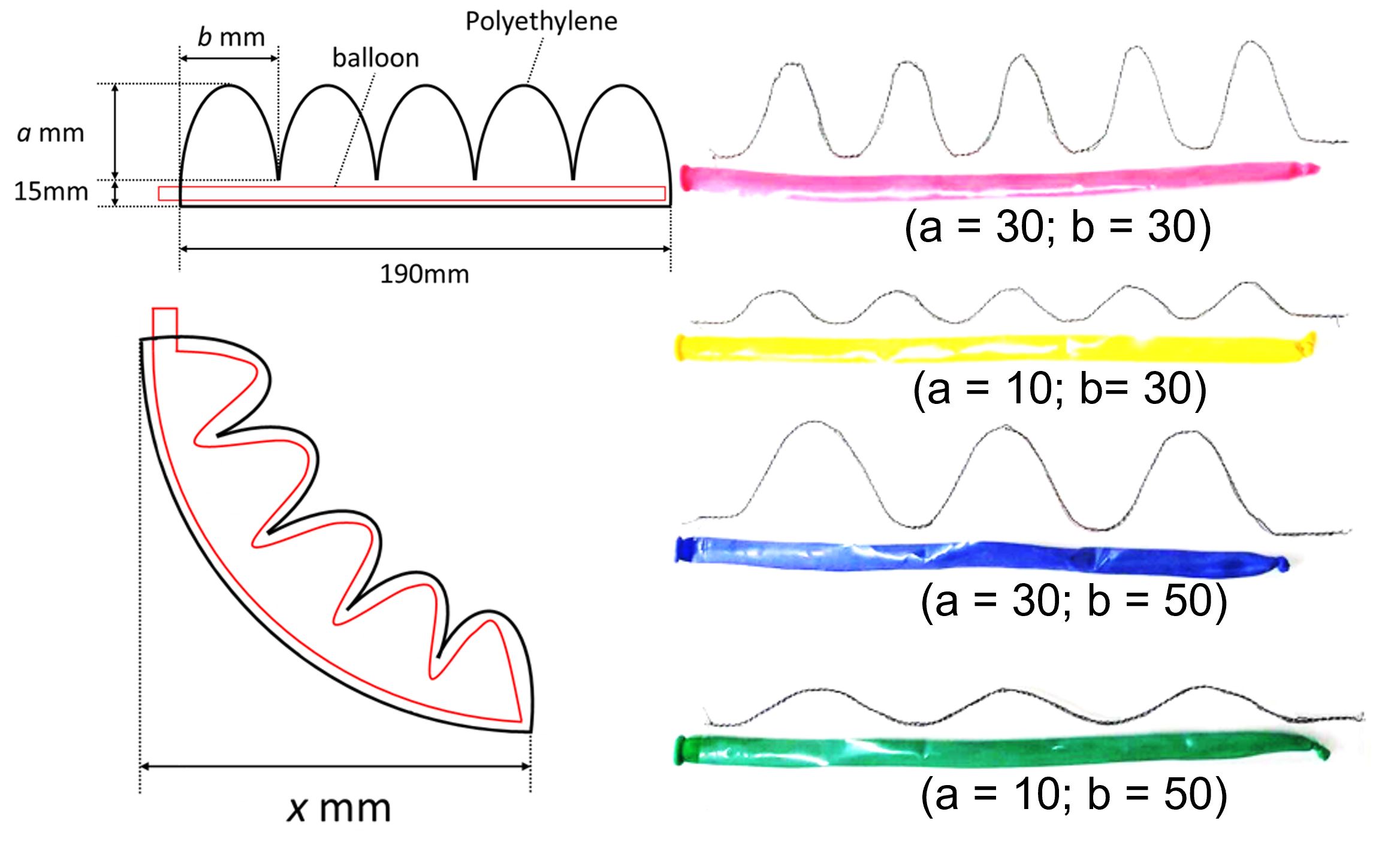}
  \caption{The configuration of xBalloon actuator and its bending state (left), and four different actuators used in performance experiments (right).}
  \label{fig:sample}
\end{figure}

In our performance experiments, four types of experimental prototypes were fabricated as shown in Figure \ref{fig:sample} (right). We defined three shape parameters for representing shape features: the height $a$ and width $b$ of one wavy shape in the actuator; the bending width $x$ after balloon inflation as shown in Figure \ref{fig:sample} (left).  We adopt the ratio $a/b$ of the height and the width of the wavy shape as the shape feature before bending. 

For simplification in the performance experiments, all the xBalloon actuators had the same length and insertion mouth size. Note that all wavy shapes in the actuator have the same shape features. The actuator is 190.0 mm and the width of the balloon insertion mouth is 15.0 mm. Four types of xBalloon actuators were fabricated with shape heights a  of 10 mm and 30 mm, and shape widths b of 30 mm and 50 mm. The bending widths $x$ after balloon inflation were measured. 

\subsubsection*{Bending Performance}
\begin{figure}[tb]
  \centering
  \includegraphics[width=\linewidth]{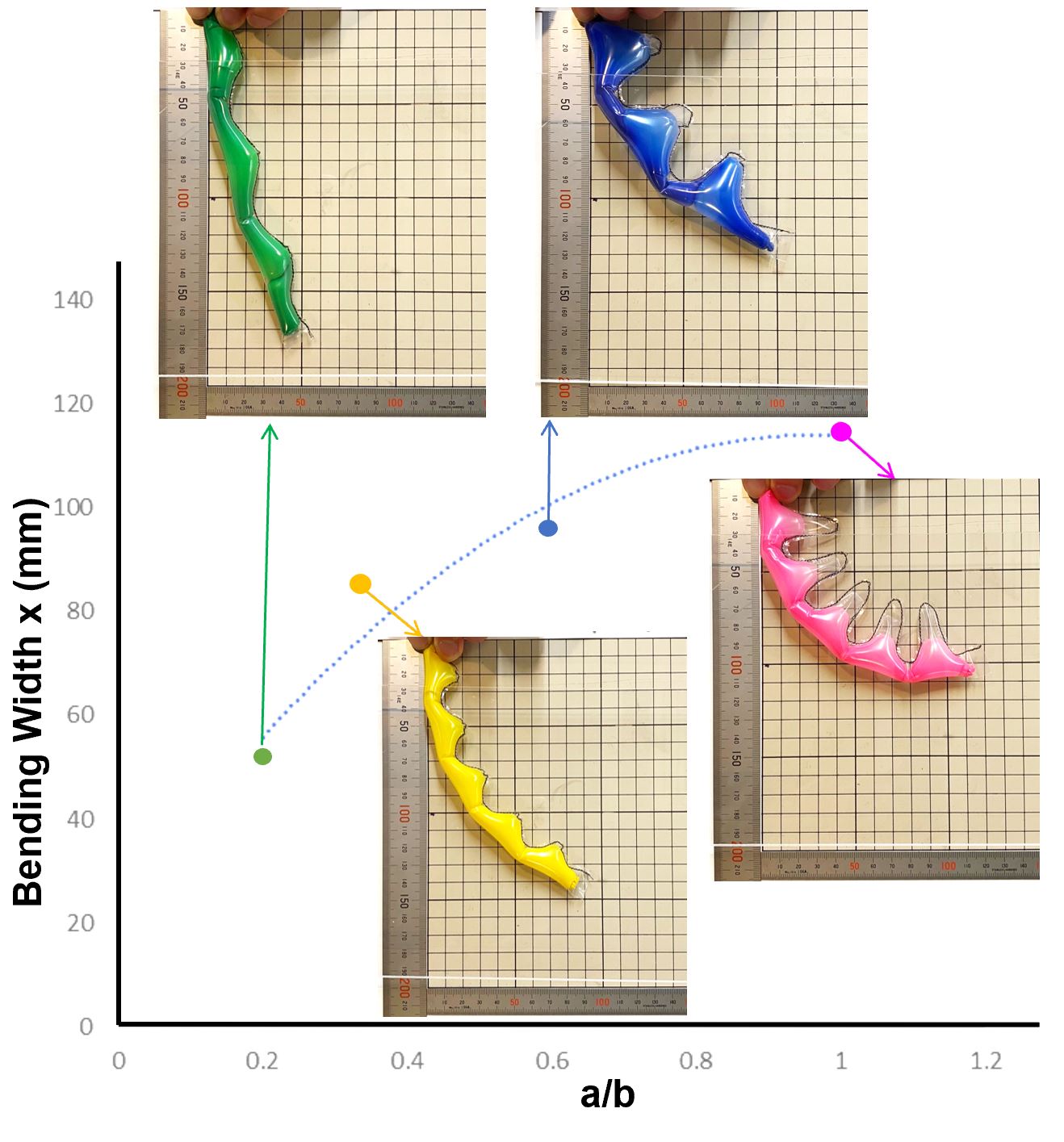}
  \caption{Relationship between the shape ratio and the bending width. Four sample actuators correspond to the same color in Figure \ref{fig:sample}.}
  \label{fig:bend}
\end{figure}

Figure \ref{fig:bend} shows the relationship between the shape ratio and the bending width. We found that the bending width $x$ [mm] increases as the value of the ratio $a/b$ between the height and width of the wavy shape increases. The bending width $x$ can be approximated with the shape ratio using the second-order polynomial approximation as follows.

\begin{equation}
x = -100.0 (\frac{a}{b})^2 + 193.0 (\frac{a}{b}) + 21.4    
\end{equation}

If all the wavy shapes are constant, the bending width can be predicted based on Equation 1 simply. In this work, this equation is adequate for simple design guideline for xBalloon actuator fabrication. Note that the accurate bending simulation of xBalloon actuator is out of the scope of this work.  

\section{User study}

\begin{figure}[tb]
  \centering
  \includegraphics[width=\linewidth]{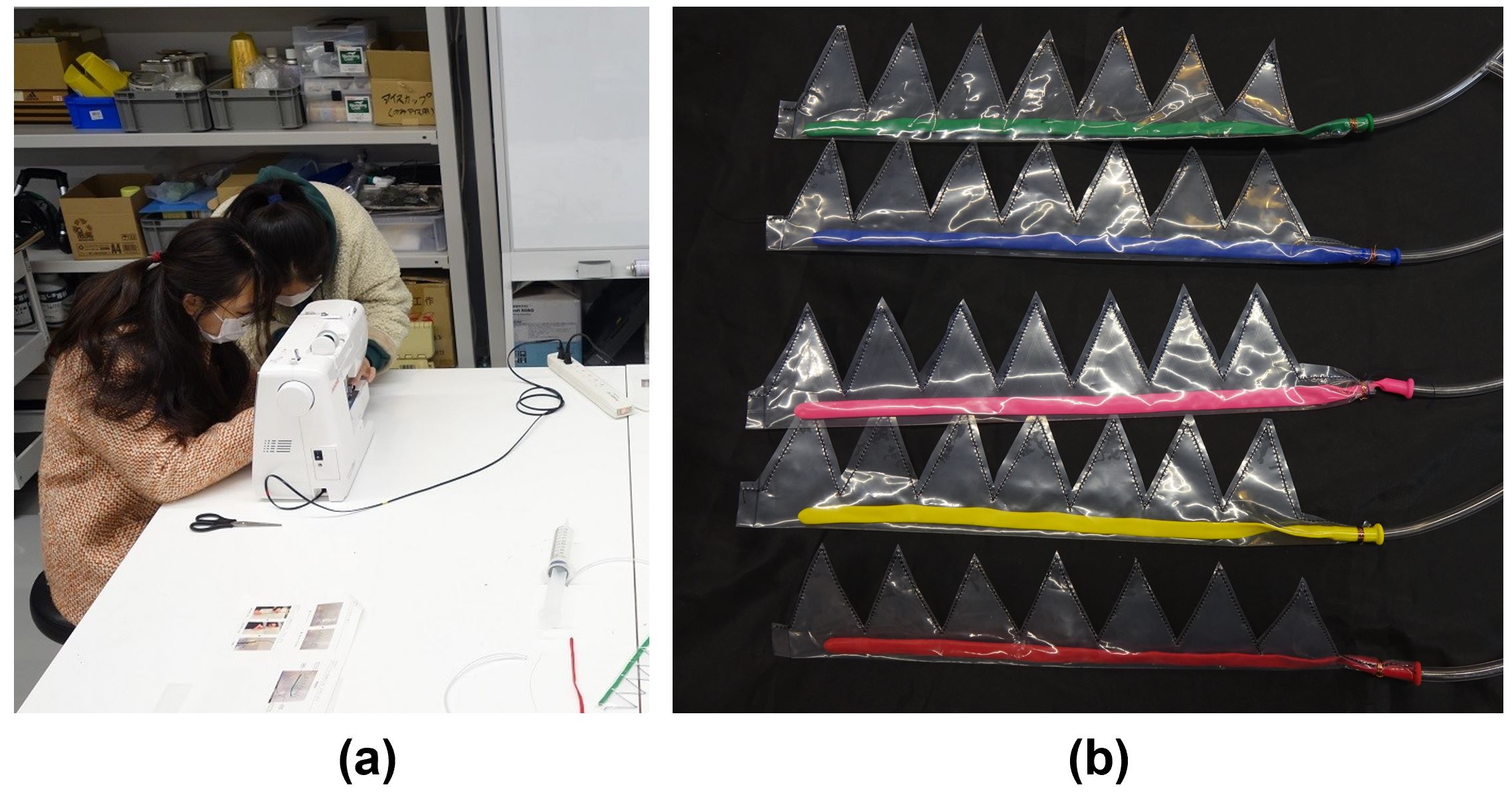}
  \caption{A snapshot of the usability experiment (a) and the fabricated xBalloon actuator from the participants (b).}
  \label{fig:exp}
\end{figure}

We conducted a usability study on the fabrication of the proposed actuator. Ten graduate students (3 males and 7 females; between 22--25 years old)  were recruited to join our user study. In this experiment, firstly the subjects were explained how to fabricate xBalloon actuator in details with its possible real-world applications, and then they were asked to actually fabricate xBalloon actuators by tracing the fabrication procedures from Step (1) to (6) in Figure \ref{fig:fab}. Both before and after experiencing the fabrication of xBalloon actuator, the subjects were asked to answer the same questionnaire survey, aiming for subjectively evaluating the fabrication procedure of xBalloon, in terms of difficulty, safety, and fun, on the 5-point scale (from 5. ``strongly agree'' to 1. ``strongly disagree''). The questionnaire survey was performed twice in order to investigate how actual fabrication experience changes their first impression for the fabrication procedure of xBalloon actuator. 
Figure \ref{fig:exp}(a) shows a snapshot of the usability experiment, and (b) shows several xBalloon actuator products made by the subjects.

The question items in the survey are listed as follows.
\begin{description}
\item[Q.1:] Do you think the fabrication process of xBalloon actuator is safe?
\item[Q.2:] Do you think you can reproduce the xBalloon actuator product of the same quality with no instructor?
\item[Q.3:] Do you think the mechanism of xBalloon actuator is interesting?
\item[Q.4:] Do you think xBalloon actuator as an actuator can work in practice?
\item[Q.5:] Do you think xBalloon actuator can be useful in various situations?
\item[Q.6:] Do you think fabricating a xBalloon actuator product is fun?
\item[Q.7:] Do you think you would like to fabricate xBalloon actuator at home?
\item[Q.8:] Do you think you would like to recommend xBalloon actuator to your family and/or friends?
\end{description}


\section{Results}

\subsubsection*{Animated Objects}
\begin{figure}[tb]
  \centering
  \includegraphics[width=\linewidth]{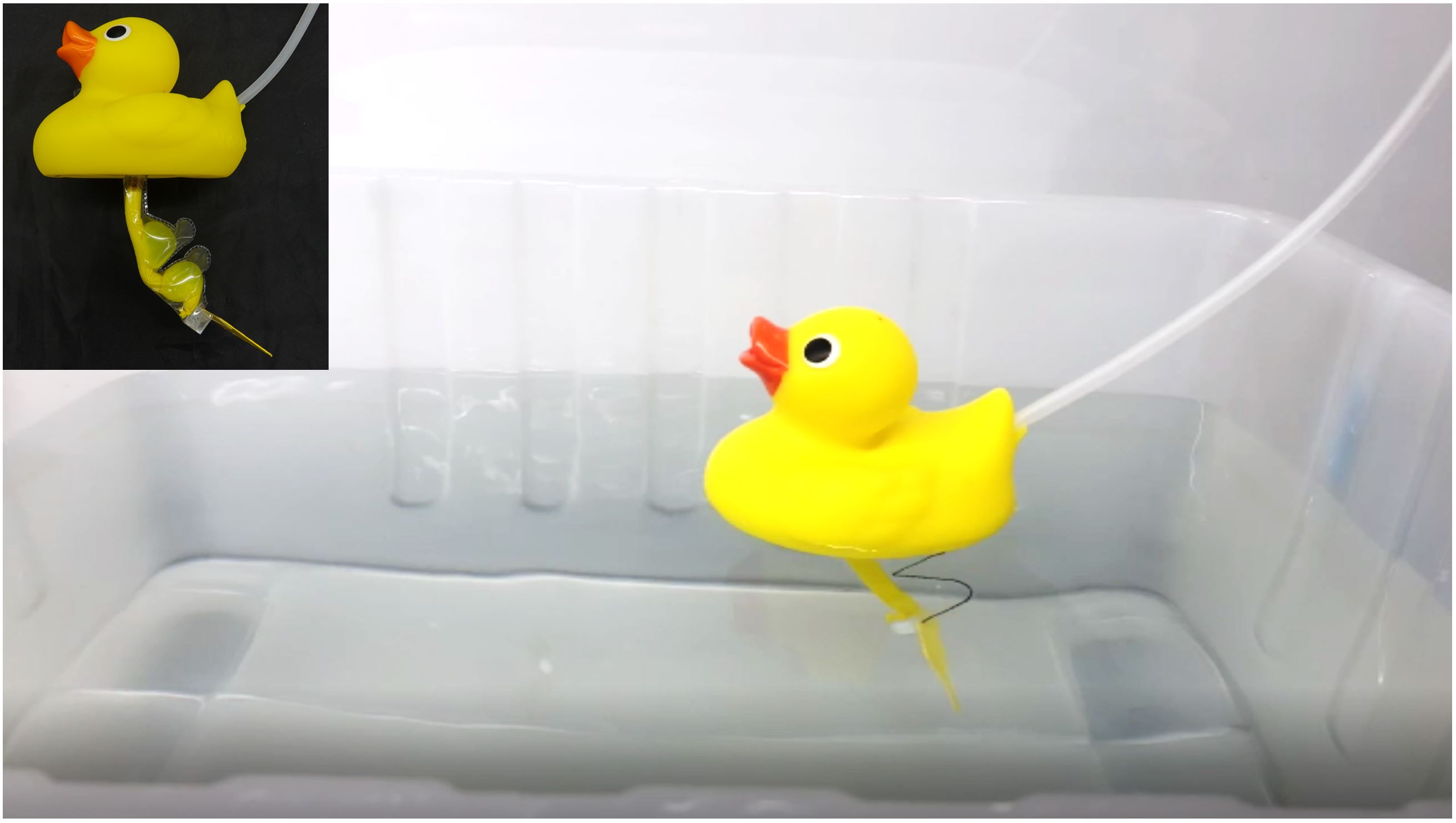}
  \caption{Swimming duck toy with one xBalloon actuator.}
  \label{fig:duck}
\end{figure}

\begin{figure}[tb]
  \centering
  \includegraphics[width=\linewidth]{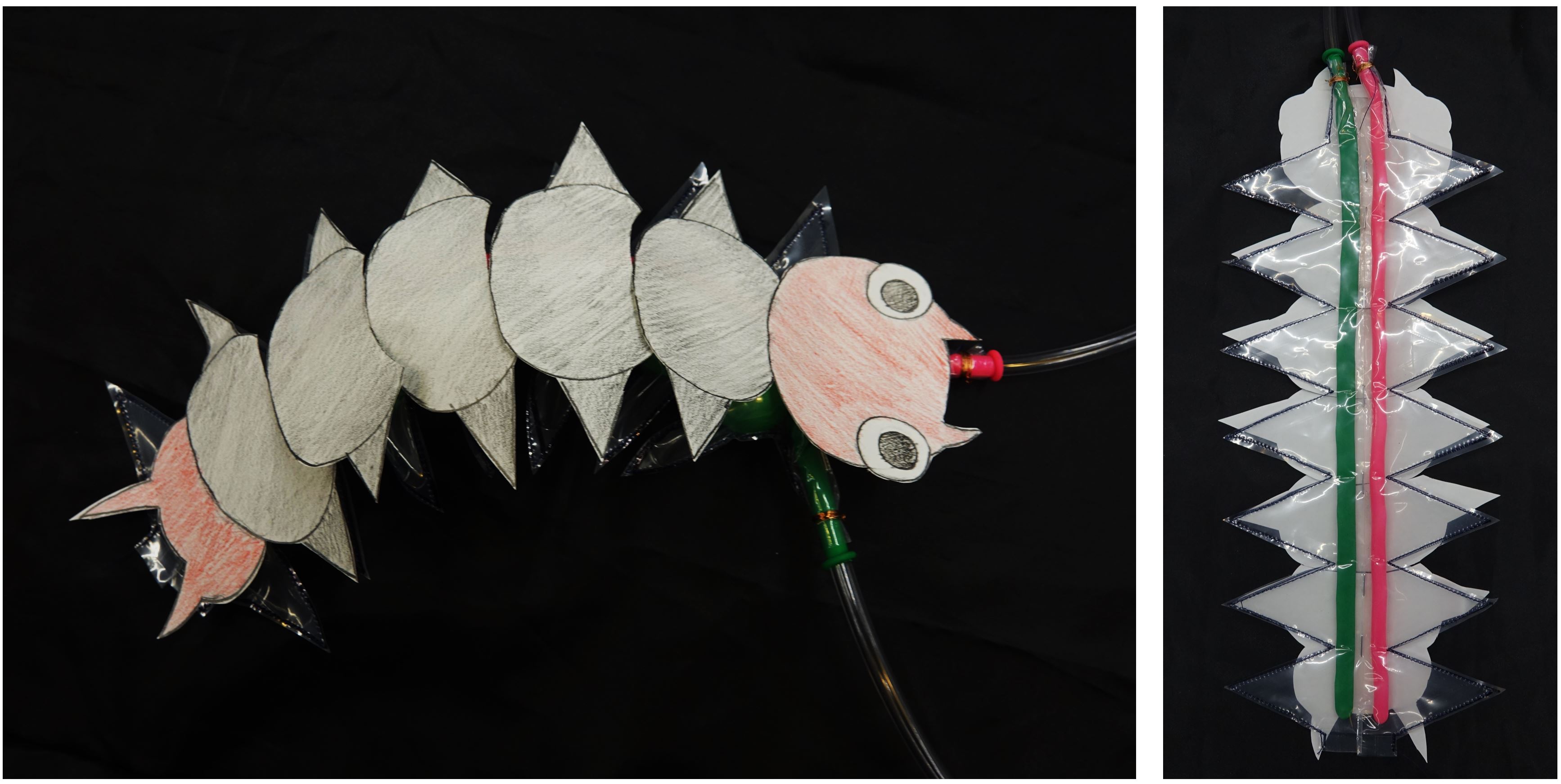}
  \caption{Walking caterpillar with with two xBalloon actuators.}
  \label{fig:cater}
\end{figure}

\begin{figure}[tb]
  \centering
  \includegraphics[width=\linewidth]{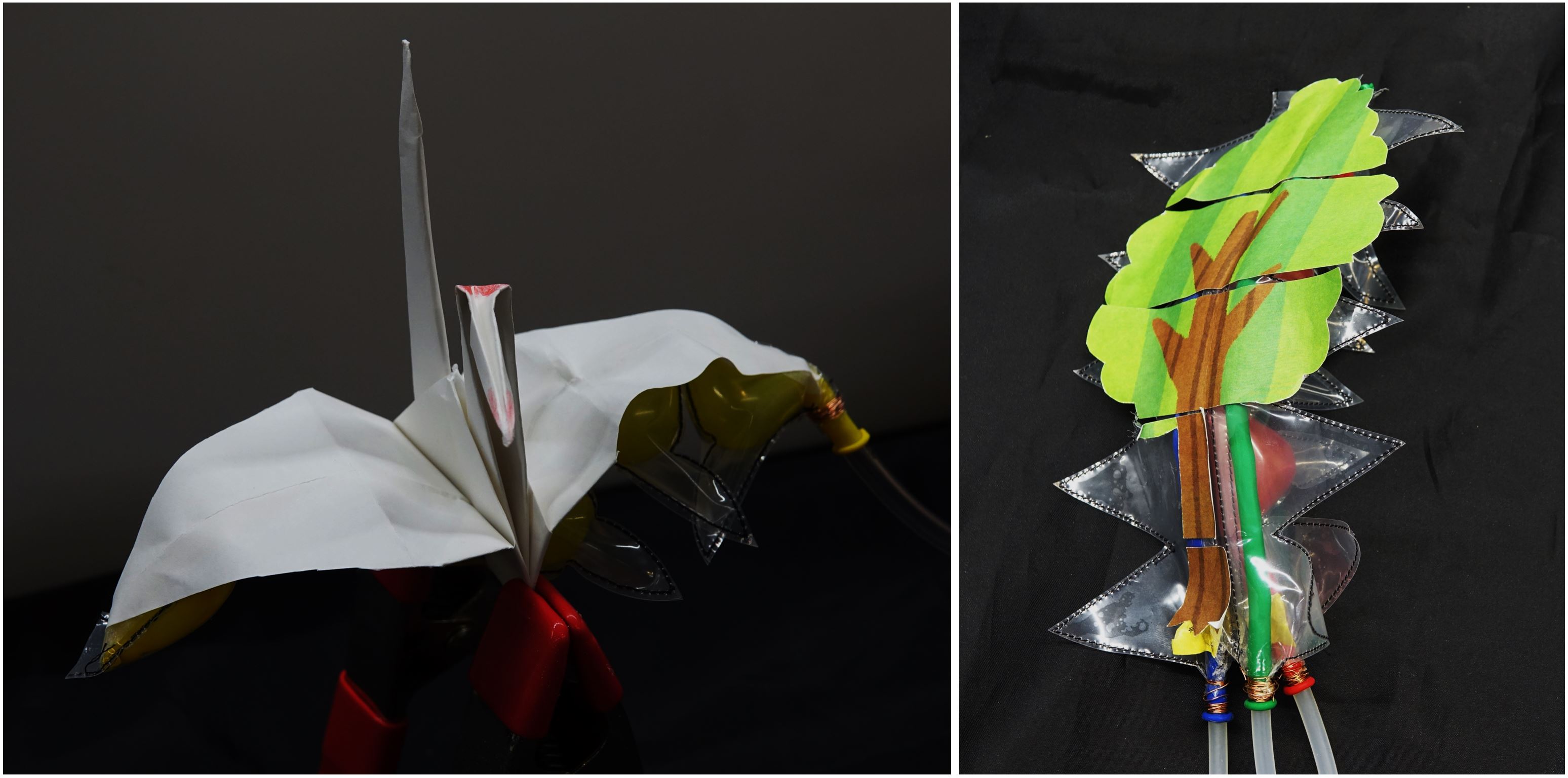}
  \caption{Flying crane with one xBalloon actuator (left), and swaying tree with three xBalloon actuators (right).}
  \label{fig:other}
\end{figure}

In this work, we consider the animated objects as the main application using the proposed xBalloon actuator. It is impressive to make static object be movable, such as toys and origami. Figure \ref{fig:duck} shows the swimming duck toy with one proposed actuator. The actuator can provide the thrust force for swimming in the water along shape bending. Figure \ref{fig:cater} shows the walking caterpillar-shape toy with two proposed actuators. The walking motions were achieved by inflating two actuators simultaneously. For other examples, a flying crane with paper origami and a swaying tree are illustrated in Figure \ref{fig:other}. For more details of these animated objects, please refer to the supplementary video.    

\subsubsection*{Questionnaire Survey}
\begin{figure}[tb]
  \centering
  \begin{tabular}{cc}
        Q.2. Reproduce & Q.3. Interesting \\
       \includegraphics[width=.48\hsize]{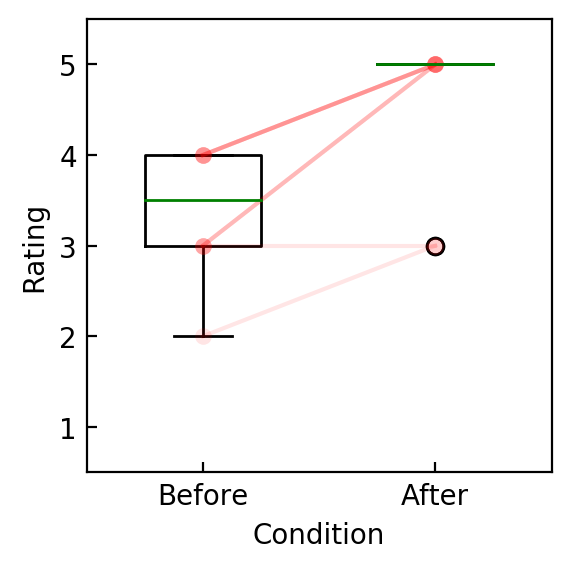} & \includegraphics[width=.48\hsize]{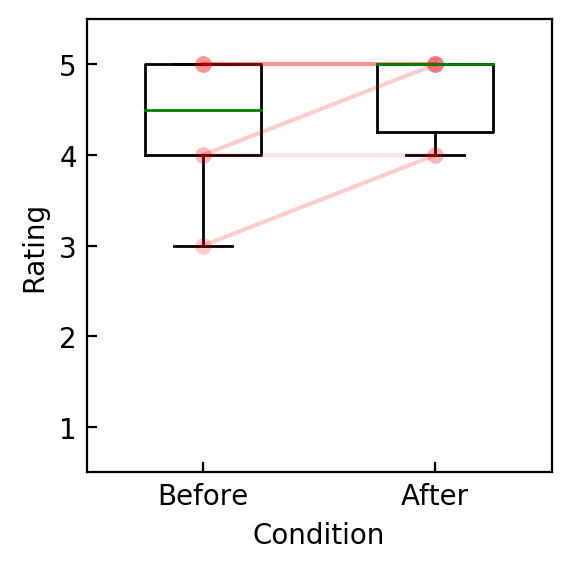} \\
        Q.6. Fun & Q.8. Recommend \\
       \includegraphics[width=.48\hsize]{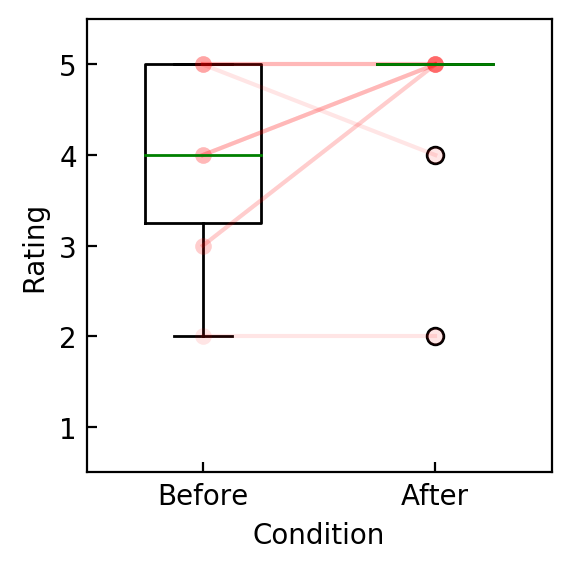} &
       \includegraphics[width=.48\hsize]{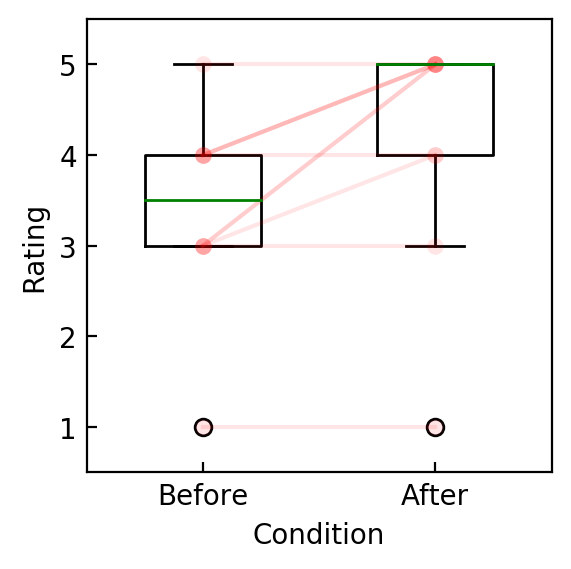} \\
  \end{tabular}
  \caption{Results in details of the before-/after-experience questionnaire surveys.}
  \label{fig:analysis}
\end{figure}

The results of the before- and after-experience questionnaire surveys are summarized in Table~\ref{tab:question}.
We observed increases in the average ratings (the Before/After columns) for all the eight question items, which suggests that actual experience of the fabrication of xBalloon actuator increases the positive impression for xBalloon actuator better than their first impression.
The proportion of non-decreasing ratings were over 70\% for all the eight question items. 
To examine whether there was a significant increase between the subjective ratings before and after the xBalloon actuator experience, the Wilcoxon's signed rank test was performed for each question item.
We found significant differences in Q.2 ($Z = 0.00$, $p < 0.01$), Q.3 ($Z = 0.00$, $p < 0.05$), Q.6 ($Z = 2.5$, $p < 0.1$), and Q.8 ($Z = 0.0$, $p < 0.05$), whose box-and-whisker plots were shown in details in Figure~\ref{fig:analysis}, in which all the pairs of the before- and after-experience ratings of the same subjects were connected with red lines.
We indeed observed that many subjects improved their impression on xBalloon actuator regarding these four question items by actually experiencing the fabrication of xBalloon actuator.


\begin{table}[tb]
  \centering
  \caption{Summary of the before-/after-experience questionnaire surveys.
  Before/After: The average ratings in the before/after conditions.
  Prop: Proportion of non-decreasing rating between the conditions.
  $Z$, $p$: The Wilcoxon's signed rank test statistic and corresponding $p$-value.}
  \label{tab:question}
  \begin{tabular}{c|ccc|cc} \hline
    Item & Before & After & Prop &  $Z$ & $p$   \\ \hline
    Q.1  & 4.3    & 4.6   &  0.9 &  2.0 & 0.256 \\
    Q.2  & 3.4    & 4.6   &  1.0 &  0.0 & 0.005 \\
    Q.3  & 4.3    & 4.7   &  1.0 &  0.0 & 0.045 \\
    Q.4  & 3.6    & 3.7   &  0.7 & 13.5 & 0.931 \\
    Q.5  & 3.6    & 4.0   &  0.8 &  6.0 & 0.317 \\
    Q.6  & 4.0    & 4.6   &  0.9 &  2.5 & 0.083 \\
    Q.7  & 3.3    & 3.8   &  0.8 &  6.0 & 0.159 \\
    Q.8  & 3.4    & 4.2   &  1.0 &  0.0 & 0.023 \\ \hline
  \end{tabular}
\end{table}

The overall comparison of the two questionnaire surveys suggests that the fabrication process of xBalloon actuator is actually easy and fun, and xBalloon actuator or its future version may have a large variety of applications in various practical situations.
Thus, the objective of this study, to propose an easy-to-fabricate inflatable structure, is considered to have been achieved.
We believe that xBalloon actuator successfully obtained such evaluation from the subjects due to the fact that the non-professional materials with no professional equipment was required in the fabrication process of xBalloon actuator.
From user experience point-of-view, no professional equipment and materials can also be important to yield quick/short-delay response or feedback to the subjects, which seem significantly to influence the subjective evaluation on `fun' throughout the long fabrication procedure.
Actually, after the usability experiment, a subject requested to take a xBalloon actuator product home to play with it at home.
This can be evidence for the easy-and-fun character of xBalloon actuator, and potential applications in an educational and entertainment context.
We are planning to hold a small-group workshop to fabricate xBalloon actuator in future.

\section{conclusion}
In this work, we proposed an inflatable balloon plastic actuator xBalloon for animated objects as novel shape-changing interfaces, which is simple and inexpensive to manufacture. The performance of the proposed xBalloon actuator was examined with bending experiment for different geometric parameters. The user experience about fabrication was verified through our user study. Additionally, multiple animated objects have been fabricated to verify the effectiveness the proposed actuator. 

In current prototype design, two main limitations of the proposed actuators are the durability and the maximum force. When the xBalloon actuator was operated many times, the film get damaged and the balloon sometimes burst. To solve these issues, we would like to try different film materials, such as cloth and feather materials. The maximum force is weak and can only be sufficient to move a light-weight objects. We think the air pressure can be enhanced with high-tension film materials. For human augmentation, we aim to embed the actuator in wearable devices such as smart clothes \cite{xie20}. In contrast to other power-based actuators, the proposed actuator can provide a safe and straightforward solution as soft actuator.  

For future work, we are planning to conduct a workshop for elementary school students using the proposed actuator for intriguing their interests in science \cite{se2020}. Besides of the application as animated objects, we would like to explore more applications in robotics and medical usages as the other soft actuators \cite{med19}. The simple fabrication approach of xBalloon can also be used as wearable devices0 with soft actuators that assist user abilities for human augmentation purpose \cite{tha2020}.

\begin{acks}
We greatly thank Genki Kakihana and Ping Zhang in research discussion and system development, and all the participants in our user study. This work was supported by JAIST Research Grant, and JSPS KAKENHI grant JP20K19845 and JP20J15087, Japan.
\end{acks}

\bibliographystyle{ACM-Reference-Format}
\bibliography{air-ref}

\end{document}